# Study of the impact of the post-MS evolution of the host star on the orbits of close-in planets ⋆,⋆⋆

## I. Sample definition and physical properties

M. I. Jones[1,2], J. S. Jenkins[1], P. Rojo[1], and C. H. F. Melo[2]

[1] Departamento de Astronomía, Universidad de Chile, Camino El Observatorio 1515, Las Condes, Santiago, Chile
[2] European Southern Observatory, Casilla 19001, Santiago, Chile

**ABSTRACT**

*Context.* To date, more than 30 planets have been discovered around giant stars, but only one of them has been found to be orbiting within 0.6 AU from the host star, in direct contrast to what is observed for FGK dwarfs. This result suggests that evolved stars destroy/engulf close-in planets during the red giant phase.
*Aims.* We are conducting a radial velocity survey of 164 bright G and K giant stars in the southern hemisphere with the aim of studying the effect of the host star evolution on the inner structure of planetary systems. In this paper we present the spectroscopic atmospheric parameters ($T_{\rm eff}$, $\log g$, $\xi$, [Fe/H]) and the physical properties (mass, radius, evolutionary status) of the program stars. In addition, rotational velocities for all of our targets were derived.
*Methods.* We used high resolution and high S/N spectra to measure the equivalent widths of many Fe I and Fe II lines, which were used to derive the atmospheric parameters by imposing local thermodynamic and ionization equilibrium. The effective temperatures and metallicities were used, along with stellar evolutionary tracks to determine the physical properties and evolutionary status of each star.
*Results.* We found that our targets are on average metal rich and they have masses between $\sim 1.0\,M_\odot$ and $3.5\,M_\odot$. In addition, we found that 122 of our targets are ascending the RGB, while 42 of them are on the HB phase.

## 1. Introduction

The determination of fundamental parameters (mass, radius, temperature, age, etc) of stars hosting exoplanets is very important, since it allows us to study how the physical properties of the host stars and the orbital parameters of the planets are related, when compared with non-planet host stars. This information can be used to test different planet formation models and to study the dynamical evolution of planetary systems. For instance, the study of the chemical abundances of stars harbouring planets led to the discovery of the planet-metallicity correlation for main-sequence stars (Gonzalez 1997; Santos et al. 2001; Fischer & Valenti 2005), which has been used as an argument in favor of the core-accretion model (Ida & Lin 2004; Alibert et al. 2005; Kennedy & Kenyon 2008).

During the main-sequence and the subgiant phase, the physical parameters of stars can be derived accurately from photometric data, since different evolutionary tracks are well separated in the color magnitude diagram for a given metallicity. However, during the red giant phase, red giant branch (RGB) and horizontal branch (HB) stars with different ages, masses and metallicities occupy a similar position in the HR diagram, making the determination of their physical parameters more difficult. In order to partially break this degeneracy, high-resolution spectra can be used to derive effective temperature ($T_{\rm eff}$), surface gravity ($\log g$), and iron abundance ([Fe/H]).

We are conducting a precision radial velocity survey of 164 bright G and K giant stars in the southern hemisphere. The main goal of this project is to determine the fraction of close-in planets (orbital periods ≲ 150 days) orbiting RGB and HB stars, and compare them in order to study how the evolution of the host star affects the inner part of planetary systems. A more detailed description of the project, along with the first results will be presented in a forthcoming paper (Jones et al. 2012, in prep.).

In this work we present the spectroscopic atmospheric parameters ($T_{\rm eff}$, $\log g$, microturbulent velocity and [Fe/H]), which are used to derive the mass, radius and evolutionary phase of the program stars. In addition, rotational velocities are measured for our targets, so can select against rapid rotators that would preclude the measurement of precise radial velocities.

The paper is organized as follows. In sections 2 and 3 we describe the targets selection, the observations and data reduction. In section 4 we present the method used to derive the atmospheric parameters. In section 5, stellar evolution models are used to derive the mass, radii and evolutionary status of the stars in our sample. In section 6 we study the dependence of macroturbulence broadening with $\log g$, which is used to derive projected rotational velocities. Finally, the summary and the discussion is presented in section 7.

---

⋆ Based on observations collected at La Silla - Paranal Observatory under programs ID's 085.C-0557 and 087.C.0476.
⋆⋆ Table 1 is available in electronic form at the CDS via anonymous ftp to cdsarc.u-strasbg.fr (130.79.128.5) or via http://cdsweb.u-strasbg.fr/cgi-bin/qcat?J/A+A/

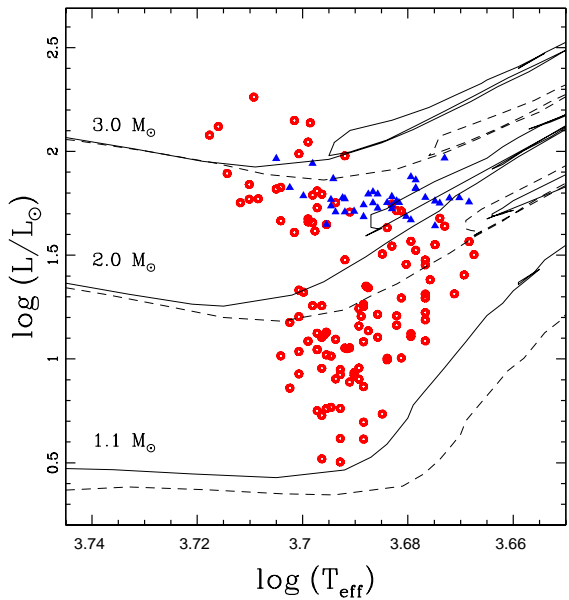

**Fig. 1.** HR diagram including all of our targets. The red open circles correspond to RGB stars, while the blue filled triangles to HB stars. Different evolutionary tracks (Salasnich et al. 2000) are overplotted for stars with $1.1\,M_\odot$, $2.0\,M_\odot$ and $3.0\,M_\odot$ (line pairs from bottom to top). The solid lines correspond to models with [Fe/H]=0.0 and the dashed lines to [Fe/H]=0.32
(A color version of this figure is available in the online journal)

## 2. Targets

We selected a total of 164 giant stars in the Southern Hemisphere from the *Hipparcos* catalog (Perryman el al. 1997) according to their position in the HR diagram ($0.8 \leq B-V \leq 1.2$; $-0.5 \leq M_V \leq 4.0$). We included stars brighter than V = 8 magnitudes, parallaxes with a precision better than 14%, and uncertainty in the Johnson B-V color less than 0.02 mags. We also removed binary systems from the sample and we add the restriction of including only those stars with a *Hipparcos* ($H_p$) photometric variability less than 0.015 mags. The *Hipparcos* ID, B-V color and V magnitudes are listed in the first three columns of Table 1. The improved *Hipparcos* parallaxes (Van Leeuwen 2007) are listed in column 4, and the uncertainties are given within brackets. We corrected the visual magnitudes using the 3D extinction maps of Arenou et al. (1992) in order to compute absolute magnitudes and luminosities (see §5). The visual extinction values ($A_V$) are listed in column 5 of Table 1.

Figure 1 shows the position of our targets in the HR diagram and their resultant evolutionary status (see §5). Evolutionary tracks from Salasnich et al. (2000) are overplotted for different stellar masses and metallicities. In summary, our sample consists of 122 RGB stars and 42 HB stars, with a range of masses between $\sim 1.0\,M_\odot$ and $3.5\,M_\odot$.

## 3. Observations and data reduction.

High resolution and high S/N spectra were taken for each of the stars in our sample. The targets were observed using the Fiber-fed Extended Range Optical Spectrograph (FEROS; Kaufer et al. 1999) mounted on the MPG/ESO 2.2 m telescope at La Silla and the Echelle Spectrograph mounted on the 1.5 m telescope at CTIO, which has now been replaced by CHIRON (Schwab et al. 2010). FEROS has a resolving power of R $\sim$ 48000, an efficiency of $\sim$ 20 % and provides an almost complete optical spectral coverage ($\sim$ 3500 - 9200 Å), which allows us to study many absorption lines in the optical range (used in the chemical analisys and in our radial velocity computations) and also to study the emission in the core of the Ca II HK lines (3933 and 3968 Å), which are used as chromospheric activity indicators (e.g. Jenkins et al. 2008 and references therein). The echelle spectrograph mounted on the 1.5 m telescope, can reach a maximum resolution of R $\sim$ 45000 with an efficiency of $\sim$ 1% and covers a spectral region between 4020 Å and 7100 Å. The exposure time of the FEROS targets (V $\leq$ 8) ranges between 60 and 480 seconds, which leads to a S/N ratio between 200-300 at 5500 Å and $\sim$ 80 at 3950 Å. The CTIO targets (V $\leq$ 6) were observed with exposure times between 180 and 300 seconds, giving rise to S/N ratios of $\sim$ 200 at 5500 Å.

The FEROS spectra were reduced in a standard fashion using the FEROS Data Reduction Software. All the calibrations (flat-fields, bias and lamps) were obtained during the afternoon, according to the standard ESO calibration plan. The reduction of CTIO spectra was performed in a similar way, using an IDL-based pipeline available for all users. As in the case of FEROS data, the calibrations were taken during the afternoon, before the nightly stellar observations.

## 4. Atmospheric parameters

We derived spectroscopic atmospheric parameters ($T_{eff}$, $\log g$, [Fe/H] and microturbulent velocity) using the equivalent widths ($EW$s) of a set of neutral and singly-ionized iron lines. We used the 2002 version of the MOOG[1] code (Sneden 1973), which solves the radiative transfer equation through a multi-layer atmospheric model by imposing excitation and ionization equilibrium (Saha-Boltzmann equation) and using the atomic parameters for each electronic transition (excitation potential ($\chi$), oscillator strength ($\log gf$) and damping constant; see § 4.1). The atmosphere models were obtained from the Kurucz (1993) grid. We linearly interpolated this grid in metallicity (fixing $T_{eff}$ and $\log g$), then in $T_{eff}$ (fixing the metallicity and $\log g$) and finally in $\log g$ to obtain the desired atmosphere model. For a detailed description of this method see Gray (2005).

### 4.1. Line list and atomic constants

We adopted the line list used in Sestito et al. (2006), which consists of a total of 159 Fe I lines and 18 Fe II lines, covering the spectral range between 5500Å and 6800Å. Features bluer than 5500Å were excluded in order to discard strongly blended lines and to avoid complications in the continuum tracing. Lines redder than 6800 Å were excluded due to the presence of many telluric features in the red part of the optical spectrum. The $\log gf$ and $\chi$ for each transition were also taken from Sestito et al. (2006; see also references therein). The collisional damping constants were computed using the Unsöld (1955) approximation, multiplied by an enhancement factor E given by: $\log E = a\chi - b$, where $a = 0.381 \pm 0.017$ and $b = 0.88 \pm 0.33$. $\chi$ corresponds to the excitation potential of the transition. This factor was derived from several Fe I features with available accurate collisional damping parameters, for stars with $T_{eff} \sim 5000\,K$ (Gratton et al.

---
[1] http://www.as.utexas.edu/ chris/moog.html

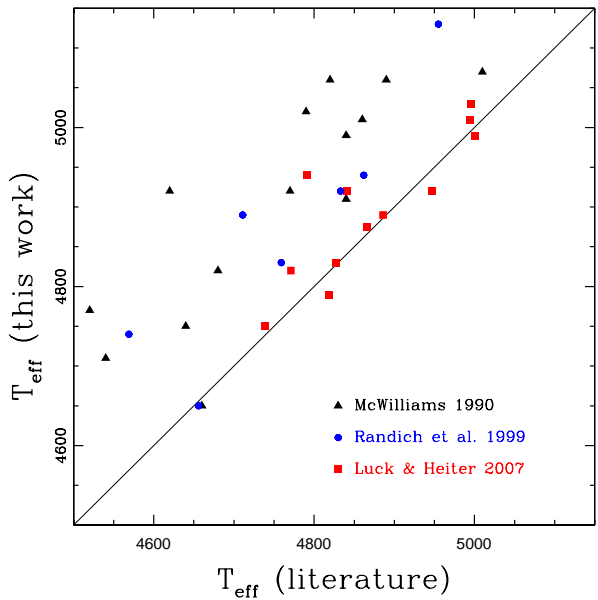

**Fig. 2.** Comparison between our derived effective temperatures with those derived by three independent works. The filled black triangles, filled red squares and filled blue circles correspond to $T_{eff}$'s derived by McWilliam (1990), Luck & Heiter (2007) and Randich et al. (1999), respectively. The solid line is the 1:1 correlation.
(A color version of this figure is available in the online journal)

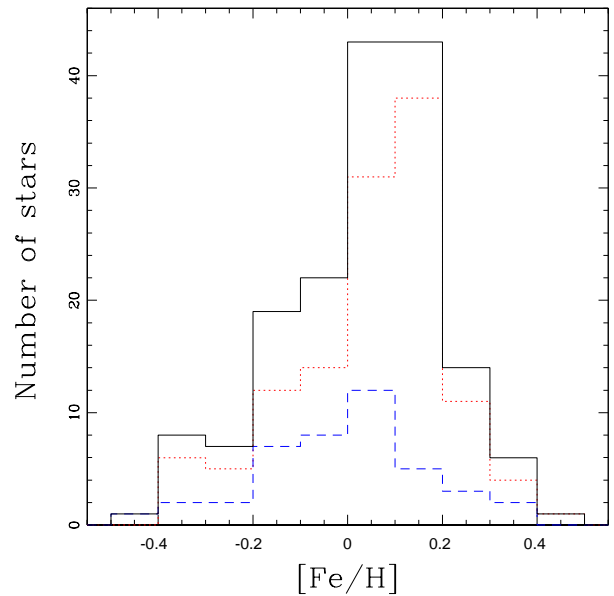

**Fig. 3.** Metallicity distribution of our targets. The red dotted line corresponds to RGB stars, while the blue dashed line to HB stars. The sum of both samples is shown with the solid black line.
(A color version of this figure is available in the online journal)

2003). Finally, we removed from the line list seven Fe I lines ($\lambda\lambda$ 5521.28, 5547.00, 5852.23, 6290.55, 6400.32, 6411.11, 6625.04 Å) and four Fe II lines ($\lambda\lambda$ 5136.80, 5525.14, 5534.85, 6383.72 Å) which were found to be the most deviant points in the abundance analysis. We examined all of these lines by eye, and we found that all of them are heavily blended leading to an inaccurate measurement of the $EW$, with the exception of the feature at 5852.23 Å, where the discrepant behavior is probably due to an error in its atomic parameters ($\log gf$ and/or $\chi$).

### 4.2. Equivalent widths

Since it is very time consuming to measure $EW$s manually, we used the code ARES[2] (Sousa et al. 2007), which computes them automatically, by applying a gaussian fit to the spectral lines. In order to test the realibility of this code we also computed some $EW$s manually, using the routine splot in IRAF[3]. We found differences < 10% between our $EW$ measurements and those derived using ARES, which are mainly due to the continuum fit. In the abundance analysis we included only those lines with $EW$ between 10 mÅ and 150 mÅ. Lines weaker than 10 mÅ were discarded because the measured equivalent width is strongly dependent on the continuum fitting. Also, lines stronger than 150 mÅ were removed from the analysis, since the gaussian fitting profile is not always appropiate to determine the $EW$ and also because these lines are more affected by collisional broadening.

### 4.3. Results and uncertainties

The effective temperatures, spectroscopic surface gravities, microturbulent velocities and iron abundances for our targets are presented in columns 6 - 9 of Table 1. We compared our resulting effective temperatures with those derived by three other independent studies that have a few stars in common with our sample. These are shown in Figure 2 where we found mean differences of $\langle \Delta T_{eff}$ (This study - McWilliam 1990)$\rangle$ = 156 ± 86 K for 14 stars in common, $\langle \Delta T_{eff}$ (This study - Randich et al. 1999)$\rangle$ = 108 ± 70 K for 6 stars in common and $\langle \Delta T_{eff}$ (This study - Luck & Heiter 2007)$\rangle$ = 24 ± 50 K for 12 stars in common. Considering all of the stars in common with these three studies we obtain $\langle \Delta T_{eff}$ (This study - literature)$\rangle$ = 98 ± 90 K. Based on this result, we adopted an uncertainty of ~ 100 K in our derived $T_{eff}$'s, which is consistent with the estimated uncertainties in similar studies (see e.g. Sestito et al. 2006, Hekker & Melendez 2007, Ghezzi et al. 2010).

Figure 3 shows a histogram of the metallicity distribution of our targets. It can be seen that most of the giant stars in our sample are metal rich, with ~ 50% having [Fe/H] between 0.0 and 0.2 dex. Also, there is no significant difference in the metallicity distribution between RGB and HB stars. The uncertainties in the metallicities were estimated from the standard deviation of individual Fe I lines in the abundance analysis, and are listed in brackets in column 9 of Table 1. These values are larger than the uncertainties in the mean, but are more realistic.

Concerning the surface gravities, several studies have revealed systematic differences between spectroscopic and photometric $\log g$'s. In the former approach, the surface gravity is varied in order to obtain the same abundance for Fe I and Fe II lines, while in the latter case the gravity of the star is derived by comparing its position in the HR diagram with theoretical evolutionary tracks. In Figure 4 we plot the difference between photometric (see § 5) and spectroscopic $\log g$'s, as a function of the effective temperature (left panel) and the mass (right panel) for our program stars. It is evident that our derived spectroscopic $\log g$'s are systematically lower (by ∼ 0.14 dex) than photometric gravities for RGB stars cooler than ∼ 5000 K (or less massive than ∼ 2.0 M$_\odot$). A similar discrepancy between photometric and spectroscopic $\log g$'s was also found by Gratton et al. (1996) and Sestito et al. (2006). However, some authors have found the oppsosite result, i.e., that the spectroscopic $\log g$'s are systematically higher than the photometric ones (e.g. Valenti & Fischer 2005, da Silva et al. 2006). This suggests that these inconsistencies in the derived surface gravities are due to systematic errors inherent to the method used for deriving iron abundances, mainly from Fe II lines, which strongly affect the final $\log g$'s (Fe I lines are quite insentive to a change in $\log g$). For instance, an overestimation/underestimation of the position of the continuum, will translate into higher/lower measured $EW$s, which are matched by a higher/lower metallicity in the curve of growth. Since the Fe II lines are on average weaker than Fe I lines, they will be more affected, hence deriving lower/higher $\log g$'s. However, we cannot discard a priori other effects like departures from LTE that might be playing a role on this (see e.g. Gratton et al. 1999).

---
[2] http://www.astro.up.pt/ sousasag/ares/
[3] IRAF is distributed by by the National Optical Astronomy Observatories, which are operated by the Association of Universities for Research in Astronomy, Inc., under cooperative agreement with the National Science Foundation.

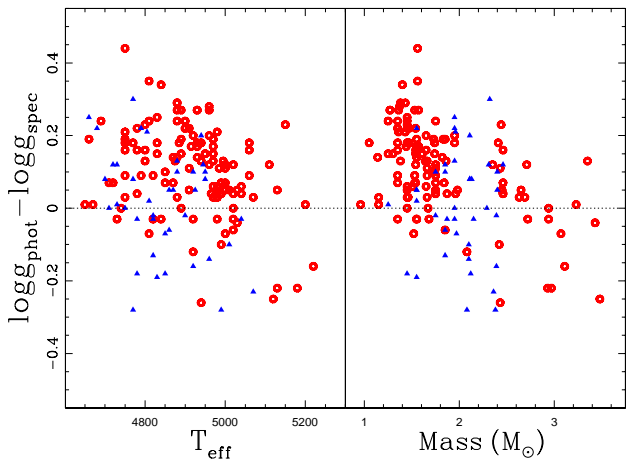

**Fig. 4.** Difference between photometric and spectroscopic $\log g$'s as a function of the effective temperature (left panel) and the mass (right panel) of our targets. The open red circles and filled blue triangles represent RGB and HB stars, respectively. The black dotted line shows the $\Delta \log g = 0.0$ boundary.
(A color version of this figure is available in the online journal)

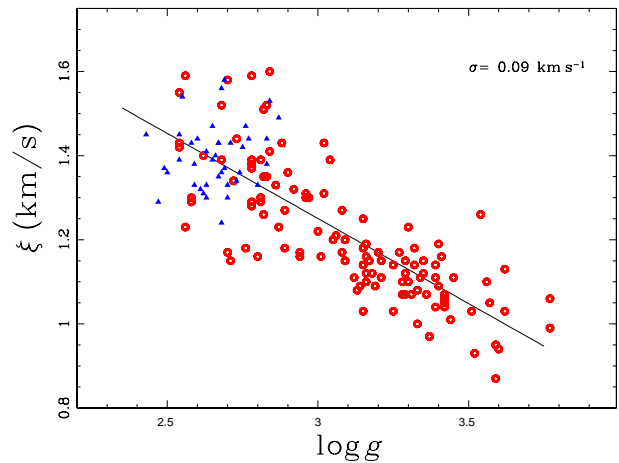

**Fig. 5.** Microturbulent velocity as a function of photometric $\log g$. The red open circles correspond to RGB stars and the blue filled triangles correspond to HB stars. A linear regression yields a fit of $\xi = 2.55 - 0.45 \log g$, with a scatter of 0.09 km s$^{-1}$.
(A color version of this figure is available in the online journal)

Finally, Figure 5 shows the microturbulent velocities as a function of the photometric $\log g$'s, which are apparently more reliable than the spectroscopic ones, as discussed above. There is a clear correlation between these two parameters, where the microturbulent velocity decreases linearly with $\log g$. Applying a linear regression yields a fit of $\xi = 2.55 - 0.45 \log g$. A similar trend was also found by other authors, who obtained: $\xi = 2.22 - 0.322 \log g$ (Gratton et al. 1996), $\xi = 1.5 - 0.13 \log g$ (Carretta et al. 2004), $\xi = 2.29 - 0.35 \log g$ (Monaco et al. 2005) and $\xi = 1.93 - 0.254 \log g$ (Marino et al. 2008).

## 5. Masses and evolutionary status

We used the evolutionary tracks from Salasnich et al. (2000) to derive the mass of each star, in a similar manner to that performed in Jenkins et al. (2009) and according to the method described the Appendix A (only available in the online version). We choose to use these models because they cover a wider range in metallicity compared to similar evolutionary tracks (for instance Girardi et al. 2002).

In addition we restricted the minimum mass of the models to 1 M$_\odot$. This is due to the fact that less massive stars spend more than ∼10 Gyr on the main-sequence, therefore nearby (d ≲ 200 pc), low-mass stars (≲ 1 M$_\odot$) are not expected to have reached the RGB phase yet. This method allows us to derive the mass of a giant star given its spectroscopic (T$_{eff}$ and [Fe/H]) and photometric ($\log L$) parameters when the evolutionary status of the star is also known. For instance, stars cooler than ∼ 5000 K and less luminous than ∼ 1.5 L$_\odot$ are ascending the RGB, and therefore their masses can be derived solely by comparing their position in the HR diagram with stellar evolution models. However, more luminous RGB stars occupy a similar position in the HR diagram as HB stars and therefore the mass and evolutionary status cannot be determined simply by comparing their effective temperature and luminosity with isomass tracks. This is the so called mass-age-metallicity degeneracy, which can be partially broken when the metallicity of the star is known (which is the case for this work).

Figure 6 shows two examples where the determination of the metallicity of the star is not enough to derive unambiguously its evolutionary status. In the upper panel we plot two models with the same metallicity (Z=0.008) but different masses and evolutionary status that cover a similar region in the HR diagram. The position of HIP 21743 (Z=0.008) is also shown. It can be seen that this star could either be ascending the RGB (somewhere in between A and B) or in the HB phase (between C and D). However, the time scale between points C and D is ∼ 5 times longer than between A and B, therefore this star is most likely to be a HB star. The lower panel shows a similar situation for HIP 68333 (Z=0.009), but this time the two models correspond to a 1.9M$_\odot$ RGB star and a 1.5M$_\odot$ HB giant. Both models have the same metallicity (Z=0.008). As in the previous case, HIP 68333 could be either somewhere in between A-B or C-D, considering the error bars in log$T_{eff}$ and LogL, but in this case the timescale between points C and D is $\sim$ 45 times longer than between A and B. Once again, based on the timescales, this giant star is most likely to be on the horizontal branch. Finally, when the position of the star in the HR diagram is very close to the zero age horizontal branch, the determination of the evolutionary status is even more complicated, since the star is as likely to be in the red giant branch as in the He-burning phase, mainly due to the uncertainty in log$T_{eff}$. The resulting masses, radii and evolutionary status of our targets are listed in columns 11, 12 and 13 of Table 1, respectively.

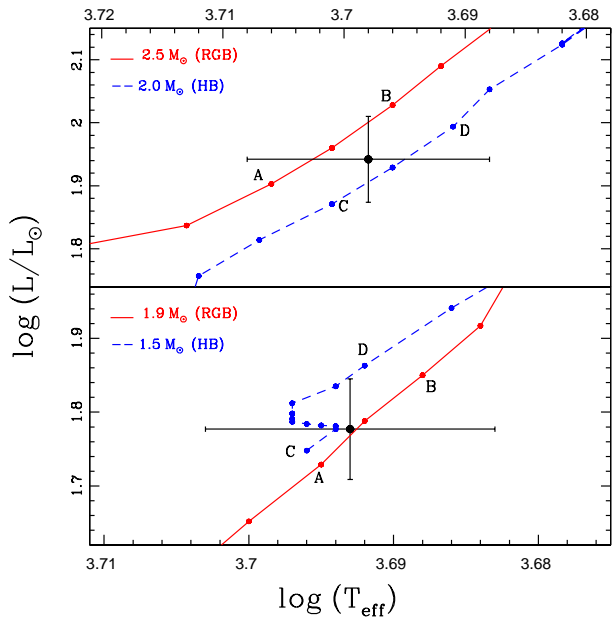

**Fig. 6.** Upper panel: Position of HIP 21743 on the HR diagram (black dot). The two closest evolutionary models from Salasnich et al. (2000) are overplotted. The red solid line corresponds to a 2.5 $M_\odot$ RGB model, while the blue dashed line to a 2.0 $M_\odot$ HB model. Both tracks have a metallicity of Z=0.008. Lower panel: same as the upper plot, but this time the masses of the models are 1.9 $M_\odot$ for an RGB star and 1.5 $M_\odot$ for a HB giant. The position of HIP 68333 is also shown.
(A color version of this figure is available in the online journal)

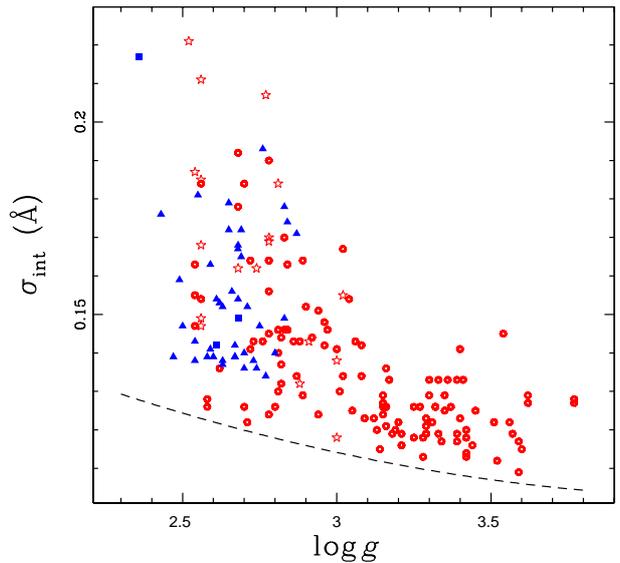

**Fig. 7.** Intrinsic broadening as a function of the photometric surface gravity. The red open circles correspond to RGB stars and the blue filled triangles to HB stars. The red open stars and blue filled squares correspond to RGB and HB stars from the literature (listed in Table 2). The macroturbulence broadening is shown by the dashed black line.
(A color version of this figure is available in the online journal)

## 6. Rotational velocities

Projected rotational velocities ($v \sin i$) were computed for the stars in our sample, according to the following procedure. First, we measured the FWHM for all of the Fe I lines used in the abundance analysis (see § 4.1 and § 4.2) between 6000 and 6100 Å and we averaged them to obtain the total FWHM of each star ($\sigma_{tot}$). The instrumental width ($\sigma_{inst}$) was measured from several ThAr lines present in the same spectral region, which is then substracted from $\sigma_{tot}$ in order to obtain the intrinsic broadening ($\sigma_{int}$) of the star by applying: $\sigma_{int} = \sqrt{\sigma_{tot}^2 - \sigma_{inst}^2}$. The second step consists of separating the contribution of the rotation and macroturbulence (non-thermal velocities), which are the two main line broadening mechanisms in giant stars.

Since we have a large dataset, it can be expected that many of our stars show projected rotational velocities close to zero (due to a low inclination, i.e., $\sin i \sim 0$) and therefore their intrinsic broadening is mainly due to macroturbulence broadening ($\sigma_{mac}$). These stars can be used to determine empirically $\sigma_{mac}$ as a function of some intrinsic property of the star (e.g. Melo et al. 2001; Jenkins et al. 2011).

In order to do this, we plotted $\sigma_{int}$ as a function of log $g$, which is shown in Figure 7. The lower envelope (dashed black line) corresponds to $\sigma_{mac}$, therefore stars lying close to this line are expected to be those with null projected rotational velocity. It can be seen that $\sigma_{mac}$ increases smoothly with decreasing log $g$, and also seems to be systematically higher for HB stars. We fit a second order polynomial of the form: $\sigma_{mac} = 0.2223 - 0.0548 \log g + 0.0063 \log g^2$, valid for RGB stars, while we just assumed a constant value for HB stars of $\sigma_{mac} = 0.134$ Å. It is worth mentioning that since the macroturbulence is as depth-dependent phenomenon (e.g. Takeda 1995), our derived $\sigma_{mac}$ is an average of the macroturbulence broadening for different lines, which are formed at different depths in the stellar atmosphere. Using the correlations derived above, we computed the rotational broadening ($\sigma_{rot}$) by applying $\sigma_{rot} = \sigma_{int} - \sigma_{mac}$.

Finally, in order to convert from $\sigma_{rot}$ to rotational velocities, we used 21 calibrators with published $v \sin i$'s derived by the Fourier transform method, which are listed in Table 2. We fit a straight line obtaining: $v \sin i = 1.18 + 42.9 \sigma_{rot}$, where $\sigma_{rot}$ is in Å and $v \sin i$ in km s$^{-1}$. The RMS of the fit is 0.89 km s$^{-1}$. We note that $v \sin i$ doesn't approach zero at $\sigma_{rot} = 0$, which produces a small systematic shift in our derived projected rotational velocities at low $\sigma_{rot}$. We applied this conversion to all of our targets to finally obtain projected rotational velocities, which are

**Table 2.** Calibrators stars

| Star | $\sigma_{rot}$ (Å) | $v \sin i$ (kms$^{-1}$) |
|---|---|---|
| HR 97 | 0.0507 | 3.9 [b] |
| HR 188 | 0.0636 | 3.0 [b] |
| HR 373 | 0.0663 | 4.5 [b] |
| HR 510 | 0.0240 | 2.9 [b] |
| HR 1030 | 0.0830 | 4.8 [b] |
| HR 1346 | 0.0429 | 2.4 [b] |
| HR 1373 | 0.0517 | 2.5 [b] |
| HR 1409 | 0.0417 | 2.5 [b] |
| HR 5516 | 0.0620 | 5.6 [b] |
| HR 5997 | 0.0450 | 3.5 [b] |
| HR 6770 | 0.0260 | 3.9 [b] |
| HR 7754 | 0.0412 | 3.2 [b] |
| HR 8093 | 0.0238 | 2.8 [b] |
| HR 8167 | 0.0885 | 5.6 [b] |
| HR 8213 | 0.0272 | 1.1 [b] |
| Pollux | 0.0157 | 2.5 [b] |
| β Crv | 0.0880 | 3.8 [a] |
| β Lep | 0.0971 | 5.1 [a] |
| β Oph | 0.0080 | 1.6 [c] |
| α Ser | 0.0620 | 5.6 [c] |
| η Ser | 0.0038 | 1.0 [c] |

**References.** (a) Gray 1982; (b) Gray 1989; (c) Carney et al. 2008

listed in the last column of Table 1.

We investigated the dependence of rotation with the luminosity and the mass of the star. Figure 8 shows our derived $v \sin i$ versus $\log L$ for all of our targets. In the lower panel the data are binned in $\Delta \log L = 0.2$ dex bins, which helps to remove the dispersion due to random inclination angles. It is clear that for RGB giants, the average rotational velocity increases smoothly with the luminosity. Also, it can be noticed in Figure 8 that even though HB stars rotate slightly slower than RGB giants having the same luminosity, the difference is not statistically significant because of the low number of HB stars in the sample (~ 20 per bin). We also plotted $v \sin i$ against the mass of our targets in Figure 9. The symbols are the same as in Figure 8, and this time we binned the data in steps of $\Delta$ mass = 0.4 M$_\odot$ bins (lower panel). It can be seen that for both, RGB and HB stars, the average rotational velocity increases with the mass of the star, and no significant difference between them is observed. Also, this plot explains the similar trend in Figure 8, since the most luminous stars in the sample are also the most massive ones.

## 7. Conclusions

High resolution and high S/N spectra were used to measure the atmospheric parameters (T$_{eff}$, $\log g$, v$_t$ and [Fe/H]) of 164 giant stars, which are the targets of our precise radial velocity program, aimed at studying the impact of the host star evolution on the inner structure of planetary systems.

We compared the resulting position in the HR diagram with evolutionary tracks in order to derive the physical properties of each star (mass, radius) and its evolutionary status. We showed the difficulties in the determination of the age and mass of stars that populate the so called "clump" in the HR diagram, since many evolutionary tracks for different masses and ages converge into this region. From the masses and radii we derived photometric gravities, which are systematically higher than the photometric values. We have also shown that the microturbulent velocity decreases linearly with $\log g$, as found previously in other works.

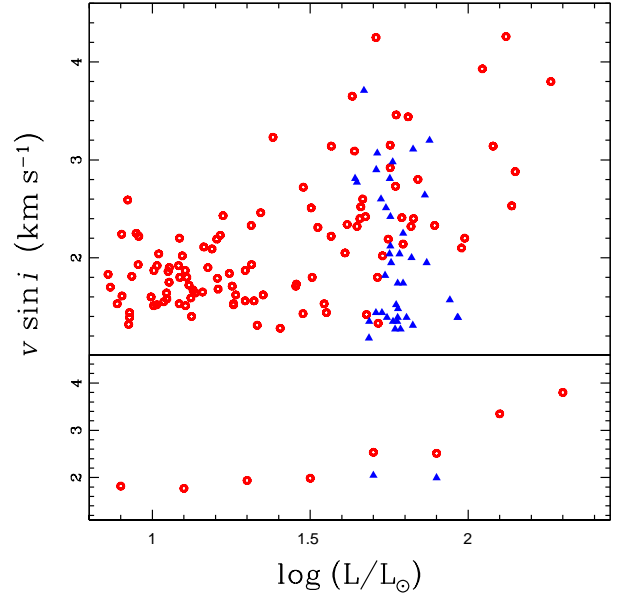

**Fig. 8.** Upper panel: Projected rotational velocities against luminosity for all of our targets. The red open circles correspond to RGB stars while the blue filled triangles to HB stars. Lower panel: Same as the upper panel, but this time the data are binned in $\Delta \log L = 0.2$ dex.
(A color version of this figure is available in the online journal)

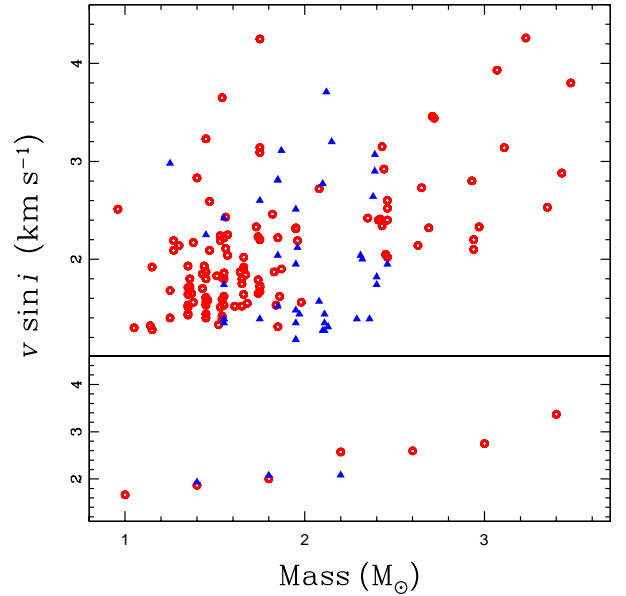

**Fig. 9.** Upper panel: Projected rotational velocities against mass for all of our targets. The red open circles correspond to RGB stars while the blue filled triangles to HB stars. Lower panel: Same as the upper panel, but this time the data are binned in $\Delta \log L = 0.2$ dex.
(A color version of this figure is available in the online journal)

Finally we computed projected rotational velocities for all of our targets. We found that the rotational velocities increase with the mass and the luminosity of the stars, and no significant difference is observed between RGB and HB stars.

*Acknowledgements.* We acknowledge to Christian Moni Bidin for his help on the iron abundances analysis and to the referee, Piercarlo Bonifacio, for his useful comments that helped to improved the quality of this work. We also thank to all of the TO's at La Silla Observatory. M. J acknowledges to ALMA funding through grant #31080027. J. S. Jenkins acknowledges funding by Fondecyt through grant 3110004 and partial support from Centro de Astrofísica FONDAP 15010003, the GEMINI-CONICYT fund and from the Comité Mixto ESO-GOBIERNO DE CHILE. P.R. acknowledges funding from FONDECYT #11080271, FONDAP and BASAL.

## Appendix A: Interpolation method

In order to determine the mass of the star we searched for the closest models (isomass tracks) in the $T_{\text{eff}}$ - $\log L$ plane by minimizing the following quantity:

$$\chi^2 = \frac{(T_{\text{eff}} - T_{\text{eff}}^M)^2}{\sigma_{T_{\text{eff}}}^2} + \frac{(L_{bol} - L_{bol}^M)^2}{\sigma_{L_{bol}}^2}, \qquad (A.1)$$

where $T_{\text{eff}}$ and $L_{bol}$ are the effective temperature and the bolometric luminosity of the star. The quantities carrying an upper script are the corresponding values of the model. The luminosities were derived using the bolometric corrections given in Alonso et al. (1999), and are listed in column 10 of Table 1. Since the evolutionary tracks are given for different metallicities, we computed a $\chi^2$ for the two set of models containing the metallicity of the star (Z), i.e., $Z_1^M \leq Z \leq Z_2^M$. Then, we used a linear weighting to determine the total $\chi^2$, by applying:

$$\chi_{tot}^2 = \alpha \chi_1^2 + \beta \chi_2^2 \qquad (A.2)$$

where $\alpha$ and $\beta$ correspond to the statistical weights given by:

$$\alpha = 1 - \frac{Z - Z_1}{Z_2 - Z_1} \qquad \beta = 1 + \frac{Z - Z_2}{Z_2 - Z_1} \qquad (A.3)$$

We repeated this procedure for the two isomass tracks that yielded the lowest total $\chi^2$ values ($\chi_{tot,1}^2$ and $\chi_{tot,2}^2$). We finally obtain the mass of the star using a linear weighting of the two closest solutions:

$$M = k_1 M_1 + k_2 M_2 \qquad (A.4)$$

where $M_1$ and $M_2$ correspond to the mass of the two isomass tracks. The weigthing constants are given by:

$$k_1 = \frac{\chi_1^2}{\chi_1^2 + \chi_2^2} \qquad k_2 = \frac{\chi_2^2}{\chi_1^2 + \chi_2^2} \qquad (A.5)$$

**Table 1.** Stellar parameter of the program stars

| HIP | B-V | V | π (mas) | $A_V$ | $T_{eff}$ (K) | log g (cm s$^{-2}$) | ξ (km s$^{-1}$) | [Fe/H] | log L ($L_\odot$) | Mass ($M_\odot$) | Radius ($R_\odot$) | Ev. Status | $v \sin i$ (km s$^{-1}$) |
|---|---|---|---|---|---|---|---|---|---|---|---|---|---|
| 242 | 0.97 | 7.76 | 7.02 (0.73) | 0.100 | 4990 | 3.14 | 1.11 | -0.04 (0.06) | 1.257 | 1.84 | 5.59 | RGB | 1.53 |
| 343 | 1.08 | 5.78 | 11.03 (0.45) | 0.100 | 4790 | 2.51 | 1.34 | 0.14 (0.10) | 1.687 | 1.95 | 10.00 | HB | 1.35 |
| 655 | 1.12 | 5.67 | 10.53 (0.38) | 0.100 | 4750 | 2.59 | 1.33 | 0.12 (0.10) | 1.778 | 1.95 | 11.75 | HB | 1.48 |
| 671 | 1.03 | 5.99 | 10.16 (0.42) | 0.100 | 4960 | 3.01 | 1.49 | -0.08 (0.16) | 1.648 | 2.10 | 8.83 | HB | 2.77 |
| 873 | 1.00 | 5.84 | 13.33 (0.35) | 0.100 | 4920 | 3.06 | 1.17 | 0.07 (0.12) | 1.478 | 2.08 | 8.11 | RGB | 2.72 |
| 1230 | 1.02 | 7.75 | 7.12 (0.92) | 0.100 | 4880 | 2.95 | 1.20 | -0.09 (0.07) | 1.264 | 1.55 | 5.89 | RGB | 1.62 |
| 1684 | 1.04 | 6.92 | 12.36 (0.58) | 0.100 | 4970 | 3.16 | 1.15 | 0.28 (0.11) | 1.104 | 1.83 | 4.74 | RGB | 1.87 |
| 1708 | 1.01 | 5.18 | 9.87 (0.41) | 0.100 | 5020 | 2.73 | 1.44 | 0.16 (0.12) | 1.989 | 2.94 | 12.28 | RGB | 2.20 |
| 3137 | 1.14 | 6.00 | 10.62 (0.43) | 0.100 | 4730 | 2.57 | 1.58 | 0.05 (0.22) | 1.642 | 1.85 | 10.20 | HB | 2.81 |
| 3436 | 1.11 | 6.01 | 17.73 (0.44) | 0.100 | 4750 | 2.90 | 1.17 | 0.17 (0.09) | 1.189 | 1.47 | 5.80 | RGB | 2.09 |
| 4293 | 1.10 | 5.45 | 14.70 (0.27) | 0.090 | 4780 | 2.74 | 1.38 | -0.07 (0.14) | 1.567 | 1.75 | 8.94 | RGB | 3.14 |
| 4587 | 0.95 | 5.62 | 10.20 (0.53) | 0.100 | 5010 | 2.80 | 1.33 | -0.17 (0.09) | 1.786 | 2.10 | 10.74 | HB | 1.27 |
| 4618 | 1.09 | 7.78 | 7.03 (0.57) | 0.123 | 4750 | 2.91 | 1.16 | 0.01 (0.09) | 1.294 | 1.45 | 6.77 | RGB | 1.56 |
| 5364 | 1.16 | 3.46 | 26.32 (0.14) | 0.100 | 4770 | 2.96 | 1.36 | 0.24 (0.19) | 1.863 | 2.38 | 11.70 | HB | 2.64 |
| 6116 | 1.03 | 7.89 | 8.06 (0.76) | 0.100 | 4850 | 3.11 | 1.12 | 0.04 (0.08) | 1.105 | 1.53 | 5.28 | RGB | 1.51 |
| 6537 | 1.07 | 3.60 | 28.66 (0.19) | 0.100 | 4820 | 2.70 | 1.56 | -0.13 (0.15) | 1.724 | 1.75 | 10.04 | HB | 2.60 |
| 7118 | 1.07 | 5.79 | 9.77 (0.41) | 0.100 | 4820 | 2.74 | 1.32 | -0.06 (0.16) | 1.783 | 1.85 | 11.18 | HB | 2.04 |
| 8541 | 1.08 | 7.88 | 5.93 (0.61) | 0.100 | 4670 | 2.70 | 1.15 | -0.15 (0.08) | 1.405 | 1.15 | 7.86 | RGB | 1.28 |
| 9313 | 1.05 | 5.57 | 11.12 (0.34) | 0.100 | 4860 | 2.71 | 1.47 | -0.03 (0.14) | 1.752 | 1.85 | 10.68 | HB | 2.81 |
| 9406 | 0.96 | 6.92 | 18.29 (0.51) | 0.100 | 4950 | 3.37 | 0.93 | -0.04 (0.06) | 0.767 | 1.25 | 3.22 | RGB | 1.40 |
| 9572 | 0.97 | 5.87 | 9.10 (0.37) | 0.100 | 5130 | 2.85 | 1.36 | 0.15 (0.10) | 1.770 | 2.65 | 9.59 | RGB | 2.73 |
| 10164 | 1.00 | 7.06 | 17.30 (0.56) | 0.100 | 4930 | 3.30 | 1.05 | 0.14 (0.09) | 0.762 | 1.36 | 3.18 | RGB | 1.72 |
| 10234 | 0.97 | 5.94 | 8.40 (0.46) | 0.092 | 4940 | 2.59 | 1.39 | -0.17 (0.06) | 1.833 | 1.85 | 11.06 | HB | 1.27 |
| 10326 | 1.01 | 5.86 | 9.40 (0.36) | 0.100 | 4950 | 2.66 | 1.36 | -0.09 (0.09) | 1.769 | 2.11 | 10.28 | HB | 1.27 |
| 10548 | 0.98 | 7.30 | 11.57 (0.51) | 0.193 | 4980 | 3.36 | 1.04 | 0.11 (0.07) | 1.045 | 1.66 | 4.32 | RGB | 1.64 |
| 11600 | 1.05 | 7.34 | 19.98 (0.76) | 0.100 | 4970 | 3.62 | 1.06 | 0.33 (0.11) | 0.519 | 1.30 | 2.47 | RGB | 2.14 |
| 11791 | 1.00 | 5.36 | 12.28 (0.45) | 0.070 | 4890 | 2.68 | 1.31 | 0.01 (0.08) | 1.733 | 2.09 | 10.72 | HB | 1.22 |
| 11867 | 1.06 | 5.91 | 8.94 (0.35) | 0.100 | 4770 | 2.32 | 1.31 | 0.18 (0.14) | 1.821 | 2.32 | 12.38 | HB | 2.00 |
| 13147 | 0.98 | 4.45 | 18.89 (0.26) | 0.100 | 4820 | 2.45 | 1.42 | -0.37 (0.08) | 1.747 | 1.53 | 11.02 | RGB | 2.19 |
| 16142 | 1.10 | 5.74 | 10.24 (0.47) | 0.124 | 4940 | 3.10 | 1.60 | 0.21 (0.21) | 1.753 | 2.43 | 9.84 | RGB | 3.15 |
| 16780 | 0.92 | 5.56 | 8.63 (0.41) | 0.144 | 5070 | 2.90 | 1.35 | -0.23 (0.07) | 1.966 | 2.36 | 11.79 | HB | 1.39 |
| 16989 | 0.98 | 5.86 | 8.90 (0.43) | 0.140 | 4960 | 2.59 | 1.35 | 0.00 (0.08) | 1.831 | 2.61 | 11.70 | RGB | 2.19 |
| 17183 | 0.95 | 6.96 | 20.82 (0.59) | 0.041 | 4930 | 3.41 | 0.87 | -0.07 (0.07) | 0.617 | 1.05 | 2.72 | RGB | 1.30 |
| 17351 | 1.19 | 4.59 | 17.70 (0.22) | 0.117 | 4700 | 2.55 | 1.33 | 0.29 (0.11) | 1.775 | 2.11 | 11.67 | HB | 1.35 |
| 17534 | 0.96 | 5.72 | 9.66 (0.32) | 0.186 | 5070 | 2.85 | 1.43 | 0.11 (0.11) | 1.819 | 2.69 | 9.88 | RGB | 2.32 |
| 17738 | 0.97 | 5.52 | 12.13 (0.30) | 0.146 | 4910 | 2.63 | 1.39 | -0.33 (0.11) | 1.708 | 1.75 | 10.04 | RGB | 4.25 |
| 18056 | 1.04 | 7.71 | 5.58 (0.78) | 0.207 | 4820 | 2.83 | 1.16 | -0.17 (0.05) | 1.544 | 1.55 | 8.23 | RGB | 1.53 |
| 18606 | 1.00 | 5.85 | 21.53 (0.41) | 0.004 | 4950 | 3.19 | 1.11 | 0.07 (0.08) | 1.014 | 1.61 | 4.50 | RGB | 1.52 |
| 19483 | 0.94 | 5.44 | 9.28 (0.37) | 0.042 | 5080 | 2.72 | 1.37 | 0.13 (0.10) | 1.907 | 2.99 | 11.70 | RGB | 3.19 |
| 19511 | 1.06 | 5.70 | 11.40 (0.51) | 0.199 | 4900 | 2.85 | 1.38 | 0.11 (0.14) | 1.712 | 2.39 | 9.87 | HB | 3.07 |
| 21154 | 1.09 | 7.42 | 10.46 (0.66) | 0.209 | 4780 | 3.00 | 1.12 | 0.18 (0.09) | 1.122 | 1.54 | 5.42 | RGB | 1.59 |
| 21685 | 1.05 | 5.46 | 16.63 (0.37) | 0.154 | 4650 | 2.55 | 1.23 | -0.31 (0.14) | 1.503 | 0.96 | 8.53 | RGB | 2.51 |
| 21743 | 0.93 | 5.56 | 9.31 (0.33) | 0.222 | 4990 | 2.82 | 1.39 | -0.36 (0.09) | 1.942 | 2.08 | 12.85 | HB | 1.57 |
| 22479 | 0.99 | 5.03 | 13.83 (0.30) | 0.176 | 4990 | 2.93 | 1.35 | 0.11 (0.10) | 1.790 | 2.42 | 9.93 | RGB | 2.41 |
| 22491 | 0.96 | 7.91 | 8.47 (0.71) | 0.230 | 5000 | 3.23 | 1.07 | -0.16 (0.06) | 1.085 | 1.45 | 4.58 | RGB | 1.53 |
| 22860 | 0.95 | 5.71 | 6.89 (0.39) | 0.242 | 5200 | 2.77 | 1.59 | 0.12 (0.07) | 2.120 | 3.23 | 12.15 | RGB | 4.26 |
| 23067 | 0.96 | 7.67 | 6.80 (0.95) | 0.138 | 5020 | 3.20 | 1.09 | -0.13 (0.06) | 1.332 | 1.85 | 6.08 | RGB | 1.31 |
| 24130 | 0.98 | 6.24 | 21.37 (0.45) | 0.049 | 4910 | 3.22 | 1.01 | -0.02 (0.07) | 0.889 | 1.35 | 3.67 | RGB | 1.53 |
| 24275 | 1.04 | 7.29 | 9.91 (0.64) | 0.099 | 4890 | 3.05 | 1.17 | 0.17 (0.10) | 1.159 | 1.74 | 5.50 | RGB | 1.65 |
| 24426 | 1.01 | 5.75 | 6.56 (0.30) | 0.187 | 5030 | 2.58 | 1.43 | 0.10 (0.10) | 2.149 | 3.43 | 16.51 | RGB | 2.88 |
| 24679 | 0.93 | 5.48 | 20.40 (0.39) | 0.051 | 4860 | 3.02 | 1.02 | -0.36 (0.07) | 1.238 | 1.08 | 5.91 | RGB | 1.51 |
| 26019 | 1.09 | 5.75 | 12.67 (0.31) | 0.124 | 4690 | 2.54 | 1.24 | 0.00 (0.10) | 1.598 | 1.63 | 9.91 | RGB | 1.83 |
| 26649 | 0.91 | 5.44 | 7.73 (0.27) | 0.114 | 5220 | 2.94 | 1.39 | 0.08 (0.10) | 2.079 | 3.11 | 11.92 | RGB | 3.14 |
| 27243 | 1.04 | 5.31 | 6.95 (0.19) | 0.183 | 5120 | 2.81 | 1.59 | 0.23 (0.12) | 2.262 | 3.48 | 16.25 | RGB | 3.80 |
| 27434 | 1.06 | 7.85 | 9.57 (0.51) | 0.152 | 4830 | 3.08 | 1.08 | 0.13 (0.09) | 0.996 | 1.45 | 4.32 | RGB | 1.60 |
| 33139 | 0.99 | 6.24 | 18.48 (0.37) | 0.100 | 5060 | 3.36 | 1.11 | 0.15 (0.09) | 1.015 | 1.66 | 4.03 | RGB | 1.92 |
| 35154 | 1.06 | 7.69 | 17.45 (0.45) | 0.104 | 4930 | 3.50 | 0.99 | 0.35 (0.12) | 0.504 | 1.27 | 2.44 | RGB | 2.19 |
| 39738 | 0.95 | 6.69 | 14.42 (0.37) | 0.062 | 4980 | 3.26 | 1.07 | -0.06 (0.07) | 1.046 | 1.46 | 4.54 | RGB | 1.58 |
| 41683 | 1.02 | 7.14 | 16.82 (0.61) | 0.110 | 4980 | 3.35 | 1.10 | 0.20 (0.09) | 0.751 | 1.43 | 3.29 | RGB | 1.85 |
| 41856 | 1.05 | 7.59 | 11.16 (0.52) | 0.100 | 4900 | 3.18 | 1.09 | 0.17 (0.09) | 0.935 | 1.54 | 4.11 | RGB | 1.81 |
| 56260 | 1.05 | 6.74 | 16.53 (0.68) | 0.019 | 4890 | 3.12 | 1.14 | 0.23 (0.12) | 0.903 | 1.53 | 4.14 | RGB | 2.24 |
| 56640 | 1.08 | 7.94 | 8.18 (0.66) | 0.163 | 4780 | 2.94 | 1.10 | 0.09 (0.09) | 1.109 | 1.45 | 5.26 | RGB | 1.80 |
| 58782 | 1.04 | 7.49 | 8.73 (0.79) | 0.115 | 4810 | 2.81 | 1.20 | -0.15 (0.08) | 1.208 | 1.25 | 5.54 | RGB | 1.68 |
| 59016 | 1.07 | 7.03 | 9.75 (0.53) | 0.104 | 4800 | 2.88 | 1.16 | 0.07 (0.10) | 1.294 | 1.64 | 6.64 | RGB | 1.87 |
| 59367 | 1.00 | 7.50 | 10.06 (0.83) | 0.019 | 4960 | 3.08 | 1.12 | -0.01 (0.07) | 1.020 | 1.57 | 4.39 | RGB | 2.04 |
| 60035 | 1.05 | 7.99 | 9.91 (0.60) | 0.293 | 4890 | 3.17 | 1.07 | 0.23 (0.11) | 0.957 | 1.55 | 4.32 | RGB | 2.22 |
| 60374 | 1.01 | 6.52 | 7.56 (0.59) | 0.131 | 4940 | 2.60 | 1.33 | -0.04 (0.08) | 1.708 | 2.11 | 9.60 | HB | 1.44 |
| 60396 | 1.06 | 6.82 | 6.66 (0.40) | 0.240 | 4810 | 2.52 | 1.45 | -0.18 (0.08) | 1.762 | 1.55 | 11.10 | HB | 1.35 |
| 62447 | 1.05 | 6.81 | 7.29 (0.46) | 0.233 | 4990 | 2.88 | 1.32 | 0.13 (0.10) | 1.657 | 2.46 | 9.025 | RGB | 2.40 |
| 63242 | 1.02 | 6.86 | 7.42 (0.49) | 0.200 | 4830 | 2.53 | 1.52 | -0.31 (0.09) | 1.633 | 1.54 | 9.41 | RGB | 3.65 |
| 63243 | 1.07 | 6.31 | 8.41 (0.38) | 0.218 | 4880 | 2.57 | 1.43 | -0.08 (0.08) | 1.743 | 1.75 | 10.15 | HB | 1.39 |
| 63583 | 1.04 | 6.65 | 7.50 (0.54) | 0.203 | 4800 | 2.46 | 1.40 | -0.22 (0.10) | 1.713 | 1.65 | 10.44 | RGB | 1.80 |
| 63981 | 1.02 | 6.75 | 9.05 (0.52) | 0.210 | 4840 | 2.85 | 1.26 | -0.22 (0.08) | 1.506 | 1.55 | 8.04 | RGB | 1.80 |
| 64580 | 1.08 | 5.91 | 9.16 (0.52) | 0.163 | 4770 | 2.55 | 1.41 | 0.02 (0.11) | 1.825 | 2.13 | 11.73 | HB | 1.31 |
| 64590 | 1.08 | 6.72 | 6.59 (0.55) | 0.222 | 4870 | 2.70 | 1.42 | 0.18 (0.12) | 1.794 | 2.40 | 10.84 | HB | 1.74 |
| 64647 | 1.09 | 7.83 | 6.70 (0.65) | 0.240 | 4870 | 2.92 | 1.21 | 0.01 (0.14) | 1.343 | 1.82 | 6.61 | RGB | 2.46 |
| 64803 | 0.96 | 5.10 | 12.66 (0.28) | 0.168 | 5060 | 2.63 | 1.39 | 0.04 (0.09) | 1.827 | 2.41 | 10.14 | RGB | 2.40 |
| 65238 | 1.08 | 7.96 | 7.28 (0.68) | 0.076 | 4810 | 2.82 | 1.15 | 0.14 (0.09) | 1.163 | 1.56 | 5.39 | RGB | 2.11 |
| 65373 | 1.03 | 6.77 | 5.17 (0.53) | 0.227 | 4920 | 2.75 | 1.34 | 0.15 (0.09) | 1.979 | 2.94 | 12.42 | RGB | 2.10 |
| 65891 | 1.00 | 6.75 | 7.35 (0.60) | 0.202 | 5000 | 2.90 | 1.30 | 0.16 (0.10) | 1.660 | 2.46 | 8.52 | RGB | 2.52 |
| 66427 | 0.94 | 5.96 | 7.86 (0.47) | 0.197 | 5180 | 3.00 | 1.37 | 0.12 (0.08) | 1.894 | 2.97 | 11.65 | RGB | 2.33 |
| 66711 | 0.99 | 7.55 | 9.64 (0.76) | 0.155 | 5000 | 3.22 | 1.00 | 0.13 (0.12) | 1.086 | 1.75 | 4.75 | RGB | 2.20 |
| 66924 | 1.02 | 5.96 | 9.06 (0.42) | 0.177 | 4860 | 2.53 | 1.43 | -0.18 (0.10) | 1.805 | 1.55 | 10.60 | HB | 1.39 |
| 67537 | 0.99 | 6.43 | 8.88 (0.46) | 0.179 | 4985 | 2.85 | 1.31 | 0.15 (0.08) | 1.617 | 2.43 | 8.57 | RGB | 2.34 |
| 67851 | 1.01 | 6.17 | 15.16 (0.39) | 0.115 | 4890 | 3.15 | 1.14 | 0.00 (0.10) | 1.244 | 1.67 | 5.71 | RGB | 1.84 |
| 67890 | 1.13 | 6.05 | 15.42 (0.43) | 0.145 | 4750 | 2.81 | 1.22 | 0.31 (0.14) | 1.313 | 1.73 | 6.90 | RGB | 2.33 |
| 68054 | 0.97 | 6.83 | 6.27 (0.51) | 0.251 | 5110 | 2.90 | 1.43 | 0.13 (0.08) | 1.772 | 2.71 | 8.44 | RGB | 3.46 |
| 68099 | 0.96 | 6.83 | 5.95 (0.55) | 0.317 | 5130 | 3.00 | 1.29 | 0.15 (0.08) | 1.841 | 2.93 | 11.57 | RGB | 2.80 |
| 68263 | 0.98 | 7.03 | 11.91 (0.88) | 0.172 | 4870 | 3.05 | 1.11 | -0.11 (0.08) | 1.136 | 1.35 | 5.31 | RGB | 1.64 |
| 68333 | 0.96 | 5.92 | 9.42 (0.48) | 0.174 | 4925 | 2.55 | 1.44 | -0.32 (0.09) | 1.777 | 1.55 | 10.36 | HB | 1.39 |
| 69065 | 1.01 | 6.39 | 8.89 (0.53) | 0.210 | 4960 | 2.62 | 1.51 | -0.22 (0.09) | 1.648 | 1.95 | 9.02 | RGB | 2.32 |
| 70261 | 1.03 | 6.80 | 7.24 (0.62) | 0.284 | 4810 | 2.65 | 1.30 | -0.38 (0.07) | 1.715 | 1.52 | 10.49 | RGB | 1.33 |
| 70514 | 1.09 | 6.83 | 9.47 (0.56) | 0.220 | 4750 | 2.65 | 1.29 | -0.12 (0.09) | 1.454 | 1.35 | 7.59 | RGB | 1.71 |
| 70987 | 1.06 | 5.99 | 10.35 (0.50) | 0.206 | 4880 | 2.64 | 1.44 | -0.03 (0.07) | 1.686 | 1.95 | 9.55 | HB | 1.18 |
| 71778 | 0.96 | 7.87 | 10.51 (0.75) | 0.110 | 5040 | 3.45 | 1.03 | 0.03 (0.12) | 0.860 | 1.51 | 3.58 | RGB | 1.83 |
| 72097 | 1.01 | 6.10 | 9.71 (0.37) | 0.193 | 5000 | 2.72 | 1.41 | -0.03 (0.10) | 1.675 | 2.35 | 9.67 | RGB | 2.42 |
| 72618 | 0.99 | 7.86 | 9.99 (1.17) | 0.115 | 4930 | 3.14 | 1.10 | -0.28 (0.11) | 0.925 | 1.14 | 4.06 | RGB | 1.32 |
| 73758 | 1.12 | 7.92 | 12.17 (0.69) | 0.095 | 4840 | 3.20 | 1.26 | 0.41 (0.15) | 0.735 | 1.40 | 3.33 | RGB | 2.83 |
| 74188 | 1.06 | 7.12 | 12.45 (0.73) | 0.185 | 4750 | 2.95 | 1.16 | 0.12 (0.12) | 1.087 | 1.36 | 5.09 | RGB | 1.80 |
| 74239 | 1.05 | 5.75 | 7.67 (0.42) | 0.256 | 5000 | 2.77 | 1.58 | 0.07 (0.15) | 2.045 | 3.07 | 13.00 | RGB | 3.93 |
| 74890 | 1.05 | 7.05 | 10.93 (0.63) | 0.195 | 4850 | 3.06 | 1.19 | 0.20 (0.13) | 1.215 | 1.74 | 5.76 | RGB | 2.23 |
| 75092 | 1.03 | 7.11 | 12.98 (0.77) | 0.364 | 4940 | 3.17 | 1.10 | 0.09 (0.11) | 1.095 | 1.66 | 4.79 | RGB | 2.02 |
| 75101 | 1.06 | 6.34 | 28.53 (0.54) | 0.079 | 4880 | 3.35 | 1.03 | 0.29 (0.13) | 0.614 | 1.27 | 2.90 | RGB | 2.09 |
| 75331 | 1.08 | 7.59 | 15.09 (0.62) | 0.152 | 4880 | 3.33 | 1.13 | 0.31 (0.14) | 0.696 | 1.38 | 3.02 | RGB | 2.17 |
| 76532 | 1.07 | 5.79 | 11.87 (0.53) | 0.394 | 4850 | 2.77 | 1.30 | 0.02 (0.08) | 1.727 | 1.97 | 10.40 | HB | 1.44 |
| 76569 | 1.08 | 5.82 | 11.45 (0.61) | 0.406 | 4830 | 2.78 | 1.38 | -0.18 (0.15) | 1.754 | 1.55 | 10.48 | HB | 2.42 |
| 77059 | 0.96 | 6.62 | 12.47 (0.47) | 0.377 | 5010 | 3.14 | 1.09 | 0.01 (0.09) | 1.322 | 1.98 | 5.93 | RGB | 1.56 |
| 77888 | 1.12 | 7.70 | 7.72 (0.65) | 0.268 | 4690 | 2.63 | 1.23 | 0.02 (0.13) | 1.314 | 1.35 | 7.08 | RGB | 1.93 |
| 78752 | 0.99 | 7.28 | 16.55 (0.92) | 0.154 | 4970 | 3.47 | 0.94 | 0.09 (0.09) | 0.728 | 1.38 | 3.09 | RGB | 1.56 |
| 78868 | 1.17 | 5.70 | 11.21 (0.36) | 0.174 | 4660 | 2.38 | 1.30 | 0.30 (0.16) | 1.756 | 1.95 | 11.22 | HB | 1.95 |
| 80672 | 1.10 | 5.79 | 9.06 (0.37) | 0.349 | 4710 | 2.47 | 1.29 | 0.03 (0.08) | 1.968 | 2.29 | 14.62 | HB | 1.39 |
| 80687 | 0.95 | 6.89 | 16.94 (0.60) | 0.332 | 5020 | 3.30 | 1.07 | 0.03 (0.07) | 0.928 | 1.54 | 4.02 | RGB | 1.39 |
| 82135 | 0.98 | 5.48 | 11.30 (0.39) | 0.196 | 4970 | 2.78 | 1.30 | 0.06 (0.09) | 1.794 | 2.63 | 10.59 | RGB | 2.14 |
| 82653 | 1.19 | 7.57 | 8.01 (0.83) | 0.618 | 4790 | 2.80 | 1.18 | 0.04 (0.10) | 1.456 | 1.75 | 7.88 | RGB | 1.73 |
| 83224 | 1.10 | 7.35 | 9.46 (0.77) | 0.288 | 4880 | 2.91 | 1.25 | 0.07 (0.08) | 1.253 | 1.75 | 5.84 | RGB | 1.71 |
| 83235 | 1.16 | 5.95 | 10.10 (0.38) | 0.218 | 4720 | 2.56 | 1.24 | 0.35 (0.13) | 1.751 | 2.31 | 11.53 | HB | 2.04 |
| 84056 | 1.03 | 6.81 | 13.31 (0.59) | 0.209 | 4960 | 3.17 | 1.12 | 0.08 (0.07) | 1.129 | 1.75 | 4.97 | RGB | 1.67 |
| 84248 | 1.07 | 5.87 | 10.00 (0.41) | 0.137 | 4730 | 2.42 | 1.45 | -0.12 (0.11) | 1.761 | 1.25 | 11.31 | HB | 2.98 |
| 85250 | 0.96 | 6.79 | 13.14 (0.54) | 0.154 | 4980 | 3.15 | 1.15 | -0.17 (0.08) | 1.124 | 1.45 | 4.96 | RGB | 1.40 |
| 86208 | 1.08 | 7.45 | 6.89 (0.59) | 0.383 | 4730 | 2.73 | 1.17 | -0.16 (0.07) | 1.551 | 1.35 | 8.61 | RGB | 1.44 |
| 86248 | 1.11 | 5.89 | 9.82 (0.43) | 0.133 | 4680 | 2.28 | 1.36 | 0.02 (0.11) | 1.777 | 1.55 | 11.62 | HB | 1.74 |
| 86368 | 1.00 | 7.43 | 13.04 (1.00) | 0.102 | 4880 | 3.13 | 1.05 | 0.12 (0.11) | 0.867 | 1.43 | 3.87 | RGB | 1.70 |
| 86786 | 0.98 | 7.21 | 10.27 (0.62) | 0.370 | 4970 | 3.07 | 1.08 | -0.12 (0.07) | 1.257 | 1.65 | 5.80 | RGB | 1.52 |
| 87273 | 1.11 | 7.02 | 11.32 (0.68) | 0.223 | 4750 | 2.64 | 1.27 | 0.21 (0.13) | 1.224 | 1.56 | 5.98 | RGB | 2.43 |

| 88684  | 0.97 | 5.74 | 27.20 (0.38) | 0.122 | 4940 | 3.24 | 1.04 | 0.04 (0.07)  | 0.904 | 1.45 | 3.90  | RGB | 1.61 |
| 90124  | 1.02 | 5.52 | 11.38 (0.35) | 0.096 | 4950 | 2.73 | 1.44 | 0.09 (0.09)  | 1.737 | 2.40 | 9.89  | HB  | 1.82 |
| 90988  | 1.04 | 7.75 | 9.17 (0.75)  | 0.136 | 4910 | 3.21 | 1.14 | 0.24 (0.14)  | 1.054 | 1.66 | 4.68  | RGB | 1.90 |
| 92367  | 0.89 | 5.80 | 9.10 (0.31)  | 0.137 | 5040 | 2.68 | 1.39 | -0.43 (0.16) | 1.826 | 1.87 | 10.74 | HB  | 3.11 |
| 95124  | 1.02 | 7.55 | 9.04 (0.61)  | 0.254 | 5040 | 3.28 | 1.18 | 0.20 (0.08)  | 1.176 | 1.87 | 4.97  | RGB | 1.90 |
| 95532  | 0.95 | 6.66 | 16.41 (0.53) | 0.082 | 4970 | 3.20 | 0.97 | -0.04 (0.08) | 0.955 | 1.44 | 4.11  | RGB | 1.93 |
| 96760  | 1.04 | 5.97 | 9.10 (0.84)  | 0.253 | 4980 | 2.86 | 1.52 | 0.12 (0.18)  | 1.810 | 2.72 | 10.53 | RGB | 3.44 |
| 97233  | 1.00 | 7.34 | 9.39 (0.70)  | 0.190 | 5020 | 3.26 | 1.23 | 0.29 (0.13)  | 1.204 | 1.96 | 5.20  | RGB | 2.19 |
| 98482  | 1.06 | 6.18 | 9.64 (0.38)  | 0.155 | 4720 | 2.51 | 1.29 | -0.17 (0.10) | 1.678 | 1.54 | 10.56 | RGB | 1.42 |
| 98575  | 0.98 | 6.01 | 9.29 (0.38)  | 0.249 | 5150 | 2.81 | 1.39 | 0.15 (0.10)  | 1.753 | 2.44 | 7.83  | RGB | 2.92 |
| 99171  | 1.02 | 5.97 | 21.30 (0.46) | 0.029 | 4830 | 3.07 | 1.03 | -0.01 (0.08) | 1.004 | 1.35 | 4.57  | RGB | 1.51 |
| 100062 | 1.00 | 5.86 | 10.31 (0.48) | 0.297 | 4920 | 2.57 | 1.43 | -0.09 (0.10) | 1.772 | 1.85 | 10.44 | HB  | 1.52 |
| 101477 | 1.00 | 5.12 | 13.94 (0.34) | 0.128 | 4980 | 2.78 | 1.35 | 0.05 (0.10)  | 1.729 | 2.46 | 10.13 | RGB | 2.02 |
| 101911 | 1.02 | 6.46 | 13.44 (0.50) | 0.045 | 4885 | 2.97 | 1.18 | 0.03 (0.08)  | 1.206 | 1.74 | 5.82  | RGB | 1.79 |
| 102014 | 1.12 | 5.47 | 13.77 (0.32) | 0.119 | 4710 | 2.65 | 1.34 | -0.02 (0.18) | 1.640 | 1.75 | 9.58  | RGB | 3.09 |
| 102773 | 1.12 | 5.41 | 10.80 (0.31) | 0.158 | 4780 | 2.58 | 1.54 | 0.01 (0.16)  | 1.878 | 2.15 | 12.92 | HB  | 3.20 |
| 103836 | 1.11 | 5.93 | 14.85 (0.49) | 0.112 | 4740 | 2.89 | 1.27 | -0.06 (0.16) | 1.382 | 1.45 | 7.17  | RGB | 3.23 |
| 104148 | 1.05 | 5.69 | 10.82 (0.64) | 0.142 | 4805 | 2.45 | 1.40 | 0.03 (0.11)  | 1.754 | 1.96 | 10.87 | HB  | 2.12 |
| 104838 | 1.01 | 6.89 | 15.67 (0.50) | 0.119 | 4900 | 3.18 | 1.06 | 0.04 (0.07)  | 0.928 | 1.45 | 3.80  | RGB | 1.44 |
| 105854 | 1.19 | 5.64 | 12.37 (0.31) | 0.162 | 4780 | 2.94 | 1.47 | 0.31 (0.18)  | 1.670 | 2.12 | 10.07 | HB  | 3.71 |
| 105856 | 1.02 | 6.70 | 14.59 (0.41) | 0.085 | 4915 | 3.08 | 1.14 | 0.09 (0.10)  | 1.050 | 1.55 | 4.90  | RGB | 1.86 |
| 106055 | 1.11 | 7.16 | 7.19 (0.75)  | 0.136 | 4770 | 2.68 | 1.33 | 0.15 (0.14)  | 1.524 | 1.95 | 8.61  | RGB | 2.31 |
| 106922 | 1.07 | 7.25 | 8.22 (0.74)  | 0.127 | 4875 | 2.96 | 1.15 | 0.12 (0.10)  | 1.351 | 1.86 | 6.45  | RGB | 1.62 |
| 107122 | 0.97 | 7.18 | 10.98 (0.81) | 0.132 | 4965 | 3.27 | 1.07 | 0.10 (0.08)  | 1.116 | 1.75 | 4.86  | RGB | 1.72 |
| 107773 | 1.02 | 5.62 | 9.65 (0.40)  | 0.167 | 4945 | 2.59 | 1.43 | 0.03 (0.10)  | 1.869 | 2.46 | 11.49 | HB  | 1.95 |
| 108543 | 1.00 | 5.50 | 7.54 (0.32)  | 0.200 | 4995 | 2.41 | 1.55 | 0.05 (0.10)  | 2.138 | 3.35 | 16.31 | RGB | 2.53 |
| 109228 | 0.95 | 7.15 | 16.48 (0.69) | 0.094 | 4960 | 3.31 | 0.95 | 0.02 (0.08)  | 0.761 | 1.37 | 3.11  | RGB | 1.65 |
| 110391 | 1.06 | 5.12 | 19.07 (0.29) | 0.088 | 4750 | 2.69 | 1.28 | -0.18 (0.08) | 1.477 | 1.35 | 7.86  | RGB | 1.43 |
| 110529 | 0.98 | 5.53 | 12.45 (0.44) | 0.161 | 5060 | 2.80 | 1.30 | 0.13 (0.11)  | 1.666 | 2.46 | 8.62  | RGB | 2.60 |
| 111515 | 0.98 | 5.97 | 10.76 (0.43) | 0.134 | 5030 | 2.97 | 1.31 | 0.14 (0.07)  | 1.610 | 2.45 | 8.03  | RGB | 2.05 |
| 111909 | 1.03 | 7.37 | 12.18 (0.74) | 0.118 | 4930 | 3.24 | 1.16 | 0.23 (0.11)  | 0.949 | 1.57 | 4.10  | RGB | 2.25 |
| 113779 | 0.97 | 7.76 | 9.21 (1.03)  | 0.147 | 5020 | 3.42 | 1.11 | 0.11 (0.08)  | 1.036 | 1.68 | 4.34  | RGB | 1.55 |
| 114408 | 0.96 | 6.47 | 15.72 (0.46) | 0.085 | 4880 | 3.12 | 1.03 | -0.25 (0.09) | 1.083 | 1.15 | 4.74  | RGB | 1.92 |
| 114775 | 1.16 | 5.77 | 13.06 (0.35) | 0.100 | 4660 | 2.57 | 1.18 | 0.26 (0.13)  | 1.566 | 1.85 | 9.41  | RGB | 2.22 |
| 114933 | 1.03 | 7.25 | 11.35 (0.93) | 0.100 | 4920 | 3.09 | 1.15 | 0.14 (0.08)  | 1.053 | 1.65 | 4.83  | RGB | 1.75 |
| 115620 | 1.06 | 5.60 | 11.23 (0.31) | 0.106 | 4820 | 2.72 | 1.37 | 0.07 (0.18)  | 1.740 | 1.95 | 10.47 | HB  | 2.51 |
| 115769 | 0.98 | 5.63 | 10.38 (0.27) | 0.110 | 4850 | 2.67 | 1.37 | -0.27 (0.13) | 1.794 | 1.45 | 11.39 | HB  | 2.25 |
| 116630 | 1.03 | 7.47 | 11.98 (0.59) | 0.102 | 4900 | 3.18 | 1.19 | 0.16 (0.14)  | 0.922 | 1.47 | 4.01  | RGB | 2.59 |
| 117314 | 1.07 | 5.74 | 10.69 (0.39) | 0.100 | 4920 | 3.00 | 1.53 | 0.07 (0.16)  | 1.709 | 2.39 | 9.75  | HB  | 2.90 |
| 117411 | 1.08 | 7.60 | 10.44 (0.71) | 0.100 | 4800 | 3.04 | 1.17 | 0.17 (0.11)  | 1.005 | 1.45 | 4.63  | RGB | 1.87 |